\documentclass[preprint]{elsarticle}

\makeatletter
\def\ps@pprintTitle{%
	\let\@oddhead\@empty
	\let\@evenhead\@empty
	\def\@oddfoot{}%
	\let\@evenfoot\@oddfoot}
\makeatother
\usepackage{graphicx}
\usepackage{natbib}
\usepackage{caption}
\usepackage{subcaption}
\usepackage{hyperref}
\usepackage[export]{adjustbox}
\usepackage{float}
\usepackage[T1]{fontenc}
\usepackage[dvipsnames]{xcolor}
\usepackage{amsmath}
\usepackage{multirow}
\usepackage{romannum}
\usepackage{enumitem}
\begin{document}

\begin{frontmatter}

\title{Random fragmentation of turbulent molecular clouds lying in the central region of giant galaxies}

\author{Suman Paul, Tanuka Chattopadhyay}
\address{Department of Applied Mathematics,\\ University of Calcutta, 92 A.P.C. Road, \\ Kolkata 700009  \\
Email: spappmath$\_$rs@caluniv.ac.in \\ \qquad \quad tchatappmath@caluniv.ac.in}

\begin{abstract}
A stochastic model of fragmentation of molecular clouds has been developed for studying the resulting Initial Mass Function (IMF) where the number of fragments, inter-occurrence time of fragmentation, masses and velocities of the fragments are random variables.  Here two turbulent patterns of the velocities of the fragments have been considered, namely, Gaussian and Gamma distributions. It is found that for Gaussian distribution of the turbulent velocity, the IMFs are shallower in general compared to Salpeter mass function. On the contrary, a skewed distribution for turbulent velocity leads to an IMF which is much closer to Salpeter mass function. The above result might be due to the fact that strong driving mechanisms e.g. shocks, arising out of a big explosion occurring at the centre of the galaxy or due to big number of supernova explosions occurring simultaneously in massive parent clouds during the evolution of star clusters embedded into them are responsible for stripping out most of the gas from the clouds. This inhibits  formation of massive stars in large numbers making the mass function a steeper one.
\end{abstract}

\begin{keyword}
Molecular clouds \sep Random fragmentation \sep Stellar mass spectrum \sep Turbulence
\end{keyword}

\end{frontmatter}


\newpage
\section{INTRODUCTION }
\label{sec:intro}
The dynamics of interstellar medium as well as the existence of baryonic dark matter (i.e the matter made of proton and neutron) is closely related to the process of star formation and galaxy evolution. Cool, dense molecular clouds which are fugacious in features, are the key ingredient relative to star formation. The fragmentation occurs due to gravitational collapse of a certain fraction of the molecular cloud when its mass is higher than a critical mass, known as Jeans mass (Jeans \cite{Jeans1902}) and this continues until the fragmented mass switches over from isothermal to adiabatic phase. Recent studies (Padoan and Nordlund \cite{Padoan2002}, \cite{Padoan2011}; Machida et al. \cite{Machida2005}; Federrath and Klessen \cite{Federrath2012}; Hopkins \cite{Hopkins2012}; Girat et al. \cite{Girart2013}) have shown that gravity and thermal pressure alone are not sufficient to explain star formation, instead turbulence, magnetic field and feedback are the other processes taking active role in star formation. The final stellar mass frequency distribution i.e the Initial Mass Function (hereafter IMF) thus obtained, delivers as the most important tool to study the star formation mechanism. However to find the origin of IMF is still difficult and one of the most profound problems in modern astrophysics.\\

Salpeter (\cite{Salpeter1955}), in his seminal work, for the first time,  showed that the IMF is a single power law distribution for the massive stars of the form,  $dN/dlogm \sim m^{-\Gamma}$, for 0.4 $M_{\odot}$ $ \le m \le$ 10 $M_{\odot}$, with index $\Gamma$ $\sim$ 1.35, where $m$ is the mass of a star, $N$ denotes the number of stars within some logarithmic  mass range $m$ and $m+dm$. In linear mass units, $dN/dm \sim m^{-\alpha}$ makes it possible to estimate the number of stars within the same mass range with $\alpha$ $=$ $\Gamma$ $+$ 1. But for the less massive stars, Miller and Scalo (\cite{Miller1979}) found that a log normal shape is more likely rather than power law shape. Few years later, the concept of segmented power law was derived (Kroupa et al. \cite{Kroupa1993}; Kroupa \cite{Kroupa2001}, \cite{Kroupa2002})  by which the shape of IMF was found to be the joining of two power law segments, one for low mass stars and another for high mass ones. Chabrier (\cite{Chabrier2003a}, \cite{Chabrier2003b}, \cite{Chabrier2005}) agreed with the lognormal shape for the low mass stars  along with a power law shape above a certain mass ($\sim$ $0.3$ $M_\odot$) (Larson \cite{Larson2005}; Chattopadhyay et al. \cite{Chattopadhyay2011}). Further theoretical and observational works were pursued in this field by many authors like De Marchi et al. (\cite{DeMarchi2005}), Zinnecker and
Yorke (\cite{Zinnecker2007}), Bastian et al. (\cite{Bastian2010})  and  Kroupa et al. (\cite{Kroupa2011}). Collecting data from two satellites Hercules and Leo \Romannum{4}, Geha et al. (\cite{Geha2013}) showed that  the IMF for the Milky Way is flatter than Kroupa IMF in the high mass regime but consistent with it in the low mass regime. By observing some massive galaxies Smith and Lucey (\cite{Smith2013}), Smith et al. (\cite{Smith2015}), Newman et al. (\cite{Newman2016}) found that these galaxies have IMF consistent with Milky Way IMF but inconsistent with Salpeter or Steeper IMF. The IMFs are most extreme in the central regions of massive galaxies (Martin-Navarro et al. \cite{Martin2015}; La Barbera et al. \cite{La2017}).\\

Arny (\cite{Arny1971}) for the first time has discussed that if the interstellar clouds are predominated by supersonic turbulence (observed by Larson \cite{Larson1981}) then the critical mass of gravitational instability is controlled by turbulent velocity rather than by the temperature. Thus the resulting mass spectrum is determined by interstellar turbulence (Larson \cite{Larson1979}, \cite{Larson1981}; Falgarone et al. \cite{Falgarone1991}). The transient structures of molecular clouds which lead to the  fragmentation process are produced mainly due to supersonic turbulent motion (McCrea \cite{McCrea1960}, \cite{McCrea1978}; Larson \cite{Larson1981}), gravitational instability, magnetic field and stellar feedback (Padoan and Nordlund \cite{Padoan2002}; Machida et al. \cite{Machida2005};  Girart et al. \cite{Girart2013}). Supersonic turbulence motion for massive stars have vital contribution to describe the theory of hierarchical fragmentation process (Mac Low and Klessen \cite{Maclow2004};   Krumholz and McKee \cite{Krumholz2005}; Hennebelle and Chabrier \cite{Hennebelle2011}; Padoan and Nordlund \cite{Padoan2011}; Federrath and Klessen \cite{Federrath2012}; Hopkins \cite{Hopkins2012}).\\

There are many models to explore the fragmentation process of massive as well as low mass stars but the actual processes are perhaps more complicated as the observational evidences are limited to certain ranges. Considering fragmentation as a random process Elmegreen and Mathieu (\cite{Elmegreen1983}) found a lognormal shape to the resulting mass spectrum after four or five iterative steps but inclusion of turbulence (Elmegreen  \cite{Elmegreen1997}) in this process changed the former shape into a power law form. Chattopadhyay et al. (\cite{Chattopadhyay2003}) considered a time dependent random fragmentation model where Poisson process is taken into account for generating the interoccurrence time between successive fragmentation events and the resulting IMF is a single power law with a steeper slope than Salpeter slope. In all the above models the resulting mass distribution is a single power law which disagrees with the observational evidences which is actually a segmented power law (Kroupa et al.  \cite{Kroupa1993}; Kroupa \cite{Kroupa2001}, \cite{Kroupa2002}). Chabrier (\cite{Chabrier2003a}, \cite{Chabrier2003b}, \cite{Chabrier2005}) also found a mass function which is a combination of a log-normal at lower masses and the Salpeter power law at higher masses. A stochastic model was developed by Chattopadhyay et al. (\cite{Chattopadhyay2011}) for hierarchical fragmentation of molecular cloud using Metropolis-Hastings algorithm which generated the fragment masses and the resulting mass spectrum as a segmented power law that is consistent with observations. The Hierarchical fragmentation process has been discussed recently by many authors (Veltchev et al. \cite{Veltchev2011}; Dobbs et al. \cite{Dobbs2014}; Heyer and Dame \cite{Heyer2015}; Contreras et al.    \cite{Contreras2016}; Li et al. \cite{Li2017}). From the above discussion it is clear that a proper theory of time dependent random fragmentation predominated by turbulence within molecular clouds may play a significant role in determining the initial mass function (IMF). The present work motivates from the above fact.\\

In the present work,  a time dependent random fragmentation model has been considered in the presence of turbulence having two distinct distributions e.g. Gaussian and Gamma, to investigate how the IMF behaves independently for the molecular cloud mass range 10$^3$ $M_\odot$ to 10$^4$ $M_\odot$ (leading to Open Clusters) and 5$\times$10$^4$ $M_\odot$ to 10$^6$ $M_\odot$ (leading to Globular Clusters) using Monte Carlo  technique. Section \ref{MATHEMATICAL MODEL} gives the description of the mathematical model. Section \ref{RESULTS AND DISCUSSIONS} describes numerical simulation, results and discussion. Finally conclusions have been outlined in Section \ref{CONCLUSION}.

\section{ MATHEMATICAL MODEL} \label{MATHEMATICAL MODEL}

\subsection{Mass distribution and fragmentation}
In the present model, a random hierarchical fragmentation of big turbulent molecular clouds has been considered where, the number of fragments ($N_F$) formed in each hierarchical step, the interoccurence time between successive fragmentation (t) in a particular time step of the hierarchy, mass of each fragment ($m_f$), number of fragmentation steps ($n$) and turbulent velocity of the fragments ($v$) are considered as random numbers. The   scenario of hierarchical fragmentation of molecular clouds has been first introduced by Hoyle (\cite{Hoyle1953}) and later density inhomogeneity has been observed in several molecular clouds (Meijerink et al. \cite{Meijerink2007}; Schleicher et al. \cite{Schleicher2010}). Subsequent observations show (Sitnik \cite{Sitnik1989}; Schwarz \cite{Schwarz1990}; Song \cite{Song2010}) there is large variation in the masses, ages, structures (continuous or filamentary), turbulence pattern in several molecular clouds observed in different regions of our Galaxy. There is no globally accepted theory so far to explain such variations. Therefore a random fragmentation model has been developed.

The concept of random fragmentation theory was 
illustrated earlier by many authors (Feller \cite{Feller1980}; 
Elmegreen and Mathieu \cite{Elmegreen1983};  Chattopadhyay et al.  \cite{Chattopadhyay2003}; Chattopadhyay et al. \cite{Chattopadhyay2011}) to find the model based stellar distribution as a result of fragmentation in a molecular cloud. Elmegreen and Mathieu (\cite{Elmegreen1983}) considered a time independent model and took Gaussian distribution for  initial distribution of fragments and studied the resulting mass spectrum using Markov Chain Monte Carlo technique. A time dependent random fragmentation model of a rod of length $l$ was considered by Feller (\cite{Feller1980}) where he computed that the probability of getting average number of random parts of a total length $l$, with each part having a length exceeding the length $x$ as,

\begin{equation} 
\label{eq1}
N_F\bigg(1-\frac{x}{l}\bigg)^{N_F-1}
\end{equation}
where, $N_F$ is the number of random parts while dividing randomly a rod of length $l$. 

Thus if $N_F$ be the total number of fragments formed within a molecular cloud during a given time interval $t_1$, the probability that the inter-occurrence time elapsed in successive fragmentations will not exceed $t$ during a particular time step $t_1$ in the hierarchy, is given by, 
\begin{equation} \label{eq2}
P(t,N_F,t_1)= 1 - \bigg(1-\frac{t}{t_1}\bigg)^{N_F},
\end{equation}
\hfill (Chattopadhyay et al. \cite{Chattopadhyay2003}),  \\ where t is a random event following the distribution function,  
 \begin{equation} \label{eq3}
P(t)= 1 - e^{-\lambda t} 
\end{equation}
\hfill (Chattopadhyay et al. \cite{Chattopadhyay2003}) \\
and the estimate of $\lambda$= $\frac{1}{\overline t}$, where, $\overline t$ $\leq$ $\frac{y}{n}$,  is the average inter-occurrence time of fragmentation (Chattopadhyay et al. \cite{Chattopadhyay2003}).  Here $y$ be the maximum duration of fragmentation in each hierarchical time step and $n$ is the number of fragmentation steps in the hierarchy. Further,  
 \begin{equation} \label{eq4}
n \leq (log M - log m_n)/log N_F,
\end{equation}

\hfill (Chattopadhyay et al. \cite{Chattopadhyay2011}),

where $M$ is the mass of the parent cloud and $m_{n}$ is the minimum mass of a fragment.

 In this work we develop a time dependent random fragmentation model with turbulence as one of the key parameters. For hierarchical fragmentation of the turbulent parent cloud, we generate the fragmented mass by considering the following expression as, 
 \begin{equation} \label{eq5}
m_f \approx \frac{10(T+100v^2)^{3/2}}{\eta^{1/2}}, 
\end{equation}

\hfill ( Fleck \cite{Robert1982}),\\
where, \\ 
$T$(in $K$) is the  temperature of the molecular cloud, \\
$\eta$ (cm$^{-3}$) is the  number density in the parent cloud, \\
$v$ (kms$^{-1}$) is the  rms turbulence velocity of a fragment \\ and 
$m_f$ ($M_\odot$) is the  mass of a clump after fragmentation.
\\
\textbf{\subsubsection{ Turbulent velocity distribution}}

Due to the  stochastic behaviour of turbulence, we choose two distribution nature  of turbulence velocity. 
  \begin{enumerate}
  \item  In the first case  we choose a Gaussian distribution for the turbulence velocity which has the following form, 
 \begin{equation} \label{eq6}
N(v)dv = \frac{1}{\sqrt{2\pi}\sigma}\thinspace e^{-\frac{(v-\mu)^2}{2\sigma^2}}dv,  - \infty < v < \infty, 
\end{equation}

 \hfill (Fleck \cite{Robert1982}; Federrath \cite{Federrath2013}), \\
 \\
 where $N(v)dv$ indicates the number density of turbulent fragments within the parent cloud with rms velocities between $v$ and $v+dv$, and $\mu$, $\sigma$ are population mean and variance. 
\\ Here estimate of $\mu$ is  $\overline{v}$ = 0.42$\times M_{eff}^{1/5}$ (Larson \cite{Larson1981}) is the average value of $v$  and 
$M_{eff} = \epsilon \times$M, where $\epsilon$ is the efficiency of star formation for the parent molecular cloud,  $M(M_\odot$) is the mass of the parent cloud at each fragmentation step and  $\sigma$ is the turbulent velocity dispersion. We set the value of velocity dispersion ($\sigma$) as equal to isothermal speed of sound ($c_s$) (Parker \cite{Parker1958}; McCrea \cite{McCrea1960}, \cite{McCrea1978})   $\sim$ 0.3 km$s^{-1}$ (Mach number is unity) (Fleck \cite{Robert1982}; Blitz \cite{Blitz1993}; Williams et al. \cite{Williams2000}) in the former case.\\
In another case we take $\sigma$ is equal to 1.5 km$s^{-1}$ (Mach number is 5) for Open clusters and 2.4 km$s^{-1}$ (Mach number is 8), 3.0 km$s^{-1}$ (Mach number is 10) for globular clusters (Larson \cite{Larson1981}; Schneider et al. \cite{Schneider2013}; Federrath et al. \cite{Federrath2016}). \\

 \item In the second case we consider Gamma distribution for turbulence which has the following form,\\
  \begin{equation} \label{eq7}
N(v) dv = \frac{\beta^\alpha v^{\alpha-1} e^{-\beta v}}{\Gamma(\alpha)} dv, v>0,\alpha,\beta>0,
\end{equation}
where, $\alpha$ and $\beta$ are shape and rate parameters respectively.
For obtaining $\alpha$ and  $\beta$ we use the following relation,  
\begin{equation} \label{eq8}
\frac{\alpha}{\beta}= A M_{eff}^{\gamma/2}, 
\end{equation}    
\hfill (Larson \cite{Larson1981}; Chieze \cite{Chieze1987}; Ossenkopf and Mac Low \cite{Ossenkopf2002}),\\
where $A$ is a random fraction and $\gamma$ = 0.3 (Ossenkopf and Mac Low \cite{Ossenkopf2002}).
\\ 
\\
The idea of non Gaussianity arises in the turbulent fluid dynamical models of dissipative system in case of large scale phenomena (Yamamoto and Kambe \cite{yamamoto1991}; Llewellyn Smith and Gille \cite{Llewellyn1998}; Pereira et al. \cite{Pereira2016}) where exponential or skewed Gamma distributions are used. Also violent phenomena involve deviation from Gaussianity e.g. high non Gaussian nature of turbulent velocity in solar flares (Jeffrey et al. \cite{Jeffrey2017}).
\end{enumerate} 

\begin{figure}
 {\includegraphics[width=\textwidth]{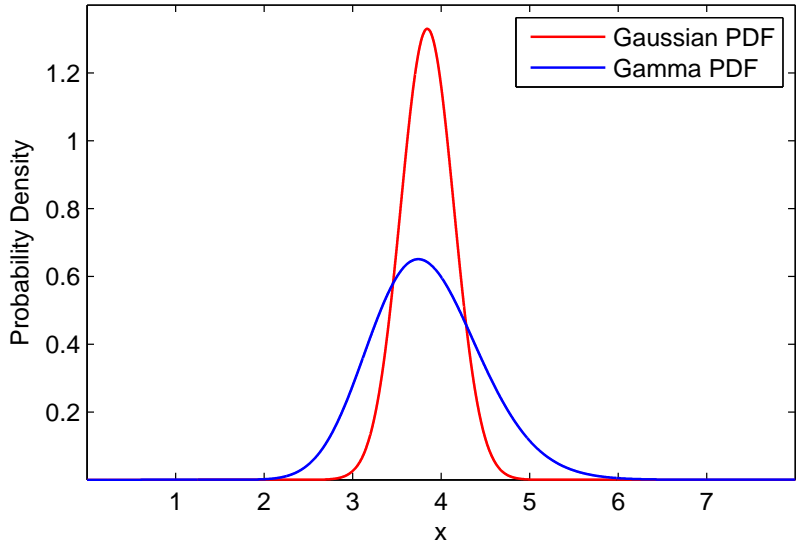}}
 \centering
\caption{Distinction between two PDFs for the turbulence velocity having  mean $=$ 3.845 and standard deviation $=$ 0.3. For Gaussian distribution $\mu$ $=$ 3.845, $\sigma$ $=$ 0.3 and for Gamma distribution shape parameter $\alpha$ $=$ 38.45 and rate parameter $\beta$ $=$ 10 i.e. mean$=$ $\frac{\alpha}{\beta}=$3.845 and standard deviation $=\frac{\alpha}{\beta^2}=$0.3}
 \label{fig:1}
 \end{figure}

\subsection{Initial choice of different parameters}   

Here we choose values of a number of parameters to generate fragment masses. The parameters involved in equations (\ref{eq2})-(\ref{eq4}) are $\lambda$, $t_1$, M, $m_n$ and n. For equation (\ref{eq5}), the corresponding parameters are temperature (T), turbulent velocity dispersion ($\sigma$) and number density of hydrogen molecules in the parent cloud ($\eta$).

\begin{table}
\centering
\caption{values of the parameters}\label{tab:1}
\begin{tabular}{@{}p{3 cm} p{4.5 cm}}
\hline 
Parameter & value  \\
\hline
$\lambda$ & $\sim$ 6.5 $\times$ 10$^{-5}$ \\
$t_1$(years) & $\sim$ 300000 \\
$M$($M_\odot$) & 10$^3-$10$^6$ \\
n   & 3 , 4 \\
$\epsilon$  & 0.1 , 0.3\\
$T$ (K)  & 10 , 20 \\
$\eta$ ($cm^{-3}$) & 10$^4$ , 5 $\times$ 10$^4$ , 10$^5$ ,  5 $\times$ 10$^5$  \\
$\sigma$ (km$s^{-1}$)& 0.3, 1.5, 2.4, 3.0 \\
$m_n$($M_\odot$) & $\sim$ 0.1\\
\hline 
\end{tabular}
\end{table}

Murray and Lin (\cite{Murray1989a}, \cite{Murray1989b}) asserted that thermal instability in a protoglobular cluster cloud is comparable to the cooling timescale $\tau_c$, where  $\tau_c$ = $\frac{3}{2}$ $\frac{kT}{\eta\Lambda(x,t)}$. For a cloud of mass 1.6 $\times$ 10$^6$ $M_\odot$ and number density $\eta$ = 270 cm$^{-3}$, $\tau_c$ = 0.9$\tau_d$, $\tau_d$ being the dynamical time. So, $\tau_c$ $<$ $\tau_d$. They also showed that the total fragmentation process held on a timescale much shorter than $\tau_d$. For a molecular cloud having number density $\eta$, varying in the range $10^4$ $cm^{-3}$ $-$ $10^6$ $cm^{-3}$ 
(Herbst and Klemperer \cite{Herbst1973}; Bally et al. \cite{Bally1987}, \cite{Bally1988}), the dynamical time scale (i.e. free fall time) is of the order $10^5$ years $-$ $10^6$ years. Hence we have chosen total time of fragmentation (viz. $t_1$) in that range (viz. Table \ref{tab:1}). In our Galaxy we have two types of star clusters , open and globular clusters in the mass regimes $10^2$ $M_\odot$ $-$ $10^4$ $M_\odot$  and $10^4$ $M_\odot$ $-$ $10^6$ $M_\odot$ respectively. Also star clusters are found to be embedded in molecular clouds (Lada and Lada  \cite{Lada2003}; Yun et al.  \cite{yun2007}). Hence the masses of the parent clouds have been considered in the range $10^3$ $M_\odot$ - $10^6$ $M_\odot$ . The number of hierarchical fragmentation steps also depends on $M$ (viz. equation (\ref{eq4})). We choose the number of fragmentation steps $n = 3$ for open clusters (Chattopadhyay et al. \cite{Chattopadhyay2003}) and $n = 4$ for globular clusters (Chattopadhyay et al. \cite{Chattopadhyay2011}) and calculate $N_F$ using equation (\ref{eq2}) while $t_1=300000$ yrs. The minimum fragment mass is taken as 0.1 $M_\odot$ ($m_n$ $\sim$ 0.1 $M_\odot$) (Haas and Anders \cite{Hass2010}).\\

\newpage
In their paper, Goldsmith and Langer (\cite{Goldsmith1978}) mentioned that in dense interstellar clouds the kinetic temperatures varied from 10 $K$ to 40 $K$. Thus temperature used for our model is taken in the above range (viz.~Table~\ref{tab:1}).

The star formation efficiency (SFE) is defined as the fraction of the stellar mass in the star-forming region and the total mass of the parent cloud. (Myers et al. \cite{Myers1986}). Federrath and Klessen (\cite{Fed2013}), suggested in their work that the range of SFE varies from 0$\%$ $-$ 20$\%$.The critical efficiency ($\epsilon$) for molecular clouds have been taken as 0.1 (Federrath and Klessen \cite{Fed2013}) and 0.3 (Lada et al. \cite{Lada1984}; Rengarajan \cite{Rengarajan1984}). All values of the parameters are listed in Table \ref{tab:1}. 

\subsection{Algorithm for numerical simulation} 
The Cumulative Distribution Function (c.d.f.) for a number of random fragments ($N_F$) is given by equation (\ref{eq2}) when the total number of fragments for a particular time interval ($t_1$) is known and the mass spectrums are generated following the various steps.\\ 

\noindent Step 1: Equation (\ref{eq3}) can be written in presence of a random fraction $k_1$ (generated at random from Uniform distribution) as,  
\begin{equation} \label{eq9}
t = - \frac{1}{\lambda}log(1-k_1)
\end{equation}
Generate one t (random) by putting the value of $\lambda$ in equation(\ref{eq9}).\\

\noindent Step 2: Now equation (\ref{eq2}) can be written in the following form as,
  \begin{equation} \label{eq10}
N_F = \frac{log(1-k_2)}{log(1-(t/t_1))}
\end{equation}
where $k_2$ is a random fraction generated from Uniform distribution. Using the above value of t, we calculate $N_F$ which is the total number of fragments after a fragmentation step.\\

\noindent Step 3: The mass spectrum can be generated using equation (\ref{eq5}) for each fragment  where the sum of the masses at each fragmentation is equal to the efficient cloud mass $\epsilon M$.\\

\noindent Step 4: Similarly we generate second generation mass spectrum for each fragment generated at random during first generation i.e. repeat steps 1 - 4 for each $n$ ( $n \sim$ 3 or 4).\\

Finally segmented power laws have been fitted for the generated fragment masses. They are shown in Figures \ref{fig:2} - \ref{fig:4} and Tables \ref{tab:2} - \ref{tab:13}. 
\\

\subsection{Robustness for the segmented power law indices $\alpha_1$ and $\alpha_2$}
For checking robustness of the results we estimate the values of the slopes for the low mass (viz. $\hat{\alpha_1}$) and high mass (viz. $\hat{\alpha_2})$ ranges following a different model developed by Chattopadhyay et al. ( \cite{Chattopadhyay2015}). 

The segmented power law  distribution for stellar masses is of the form, 
\begin{equation} \label{eq11}
\frac{dN}{dm} = 
\begin{cases}
Am^{-\alpha_1}, \quad  \quad {m_{min}\le m \le m_c} \\
Bm^{-\alpha_2}, \quad \quad {m_c \le m \le m_{max}}
\end{cases}
\end{equation}
where, $m_{min}$ and $m_{max}$ are the minimum and maximum masses of the stars, $m_c$ is the characteristic mass at which the turnover occurs, $A$, $B$, $\alpha_1$ and $\alpha_2$  are constants.\\
\\
For cross-checking  we estimate $\alpha_1$ in the low mass regime for our generated masses of the fragments, using the following equation as, 
 \begin{equation} \label{eq12}
m_{f_1}^{1- \hat{\alpha_1}}= r_1 m_c^{1- \hat{\alpha_1}}+(1-r_1) m_{min}^{1-\hat{\alpha_1}}, 
\end{equation}
\hfill (Chattopadhyay et al. \cite{Chattopadhyay2015}) \\
Thus,  for $r_1$ = 0, $m_{f_1}$ = $m_{min}$ and  for  $r_1$ = 1, $m_{f_1}$ = $m_{max}$, 
where, $m_{f_1}$ are the masses generated from equation (\ref{eq5}) between $m_{min}$,  $m_c$ and $r_1$ is a  random fraction generated between 0 and 1.\\
\\
Similarly we estimate $\alpha_2$ in the high mass regime using the following equation, 
 \begin{equation} \label{eq13}
m_{f_2}^{1-\hat{\alpha_2}}= r_2 m_{max}^{1-\hat{\alpha_2}}+(1-r_2) m_c^{1-\hat{\alpha_2}}, 
\end{equation}
\hfill (Chattopadhyay et al. \cite{Chattopadhyay2015}) \\
Here, when  $r_2$ = 0, $m_{f_2}$ = $m_c$ and  when $r_2$ = 1, $m_{f_2}$ = $m_{max}$
where, $m_{f_2}$ are the masses generated from equation (\ref{eq5}) between $m_c$ and $m_{max}$ and $r_2$ is a random fraction between 0 and 1.

The estimated values (viz. $\hat{\alpha_1}$ and $\hat{\alpha_2}$) are shown in Tables \ref{tab:4} - \ref{tab:5}, Tables \ref{tab:8} - \ref{tab:9} and Tables \ref{tab:12} - \ref{tab:13} (columns 7 - 8 of Tables \ref{tab:4}- \ref{tab:5}, \ref{tab:8} - \ref{tab:9}, \ref{tab:12} - \ref{tab:13}  respectively).

\section{RESULTS AND DISCUSSIONS} \label{RESULTS AND DISCUSSIONS}
\begin{figure}
  \fbox{\includegraphics[width=1.1\textwidth]{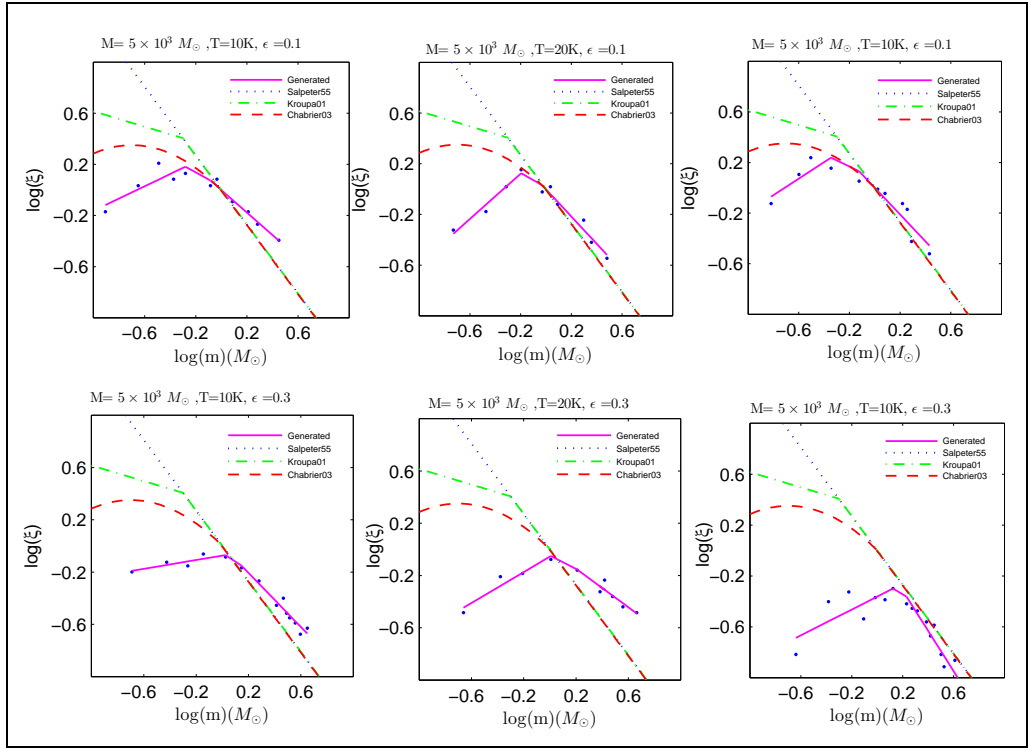}}
  \centering
\caption{Segmented power law fit for Gaussian distribution of turbulent velocity with $\sigma=$0.3 kms$^{-1}$ (1st column), $\sigma=$1.5 kms$^{-1}$ (2nd column) and segmented power law fit for Gamma distribution of turbulent velocity (3rd column) and compared with the observed IMF in the Milky Way by Salpeter (\cite{Salpeter1955}), Kroupa (\cite{Kroupa2001}) and Chabrier (\cite{Chabrier2003a})}
  \label{fig:2}
  \end{figure}

\begin{figure}
  \fbox{\includegraphics[width=1.1\textwidth]{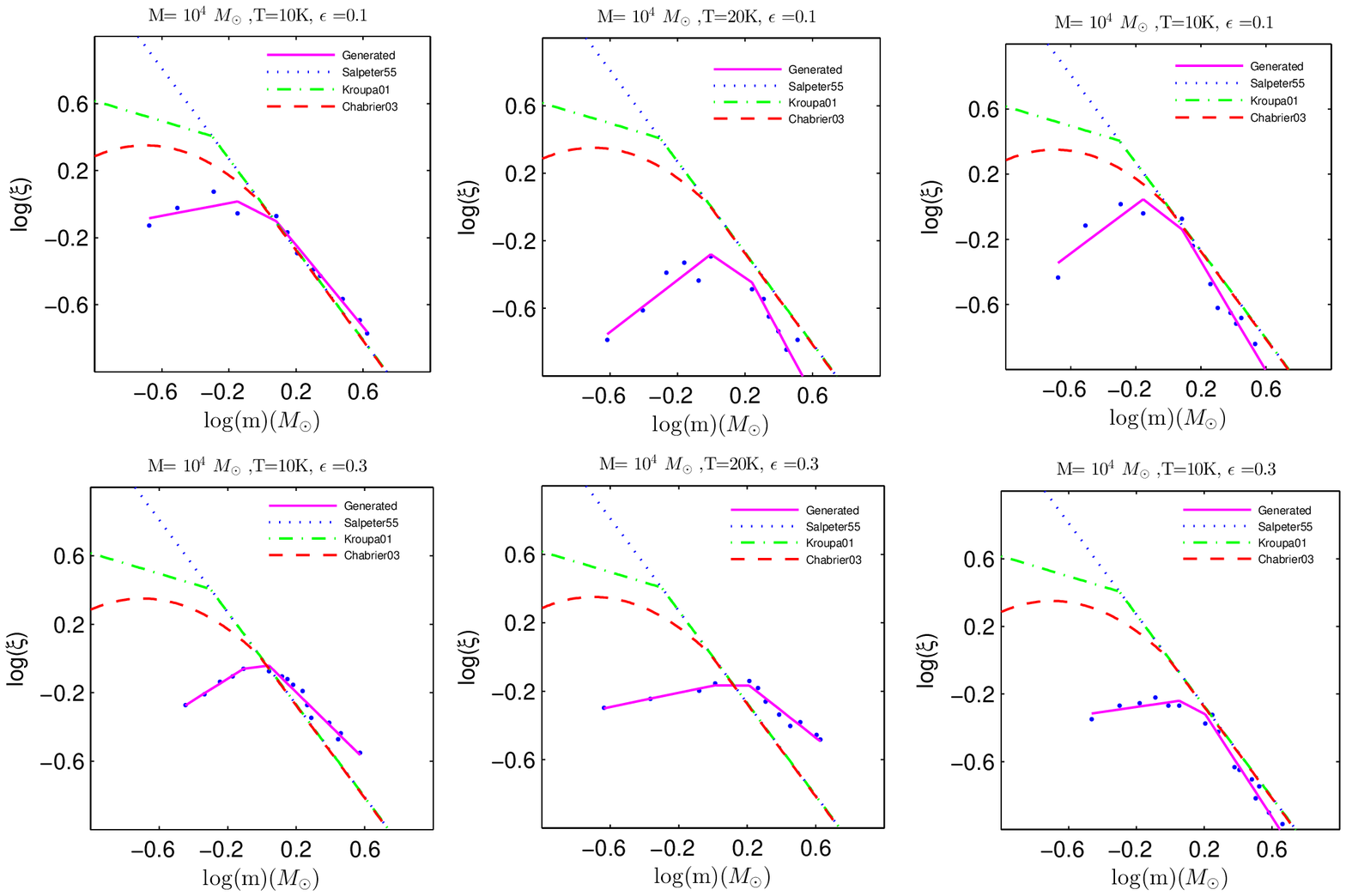}}
  \centering
\caption{Segmented power law fit for Gaussian distribution of turbulent velocity with $\sigma=$0.3 kms$^{-1}$ (1st column), $\sigma=$1.5 kms$^{-1}$ (2nd column) and segmented power law fit for Gamma distribution of turbulent velocity (3rd column) and compared with the observed IMF in the Milky Way by Salpeter (\cite{Salpeter1955}), Kroupa (\cite{Kroupa2001}) and Chabrier (\cite{Chabrier2003a})}
  \label{fig:3}
  \end{figure}

\begin{figure}
 \fbox{\includegraphics[width=1.1\textwidth]{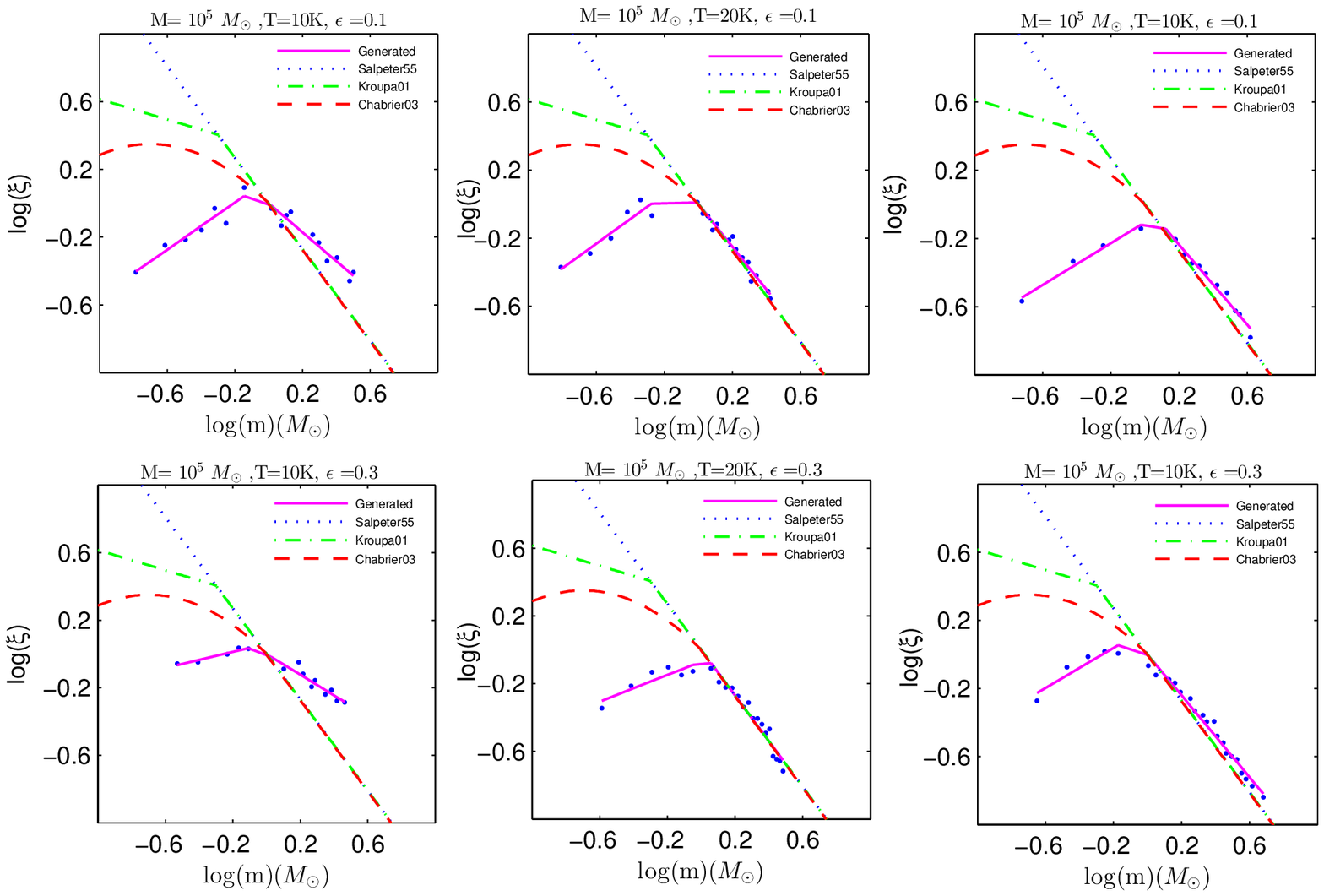}}
 \centering
\caption{Segmented power law fit for Gaussian distribution of turbulent velocity with $\sigma=$0.3 kms$^{-1}$ (1st column), $\sigma=$2.4 kms$^{-1}$ (2nd column) and segmented power law fit for Gamma distribution of turbulent velocity (3rd column) and compared with the observed IMF in the Milky Way by Salpeter (\cite{Salpeter1955}), Kroupa (\cite{Kroupa2001}) and Chabrier (\cite{Chabrier2003a})}
 \label{fig:4}
 \end{figure}

 In the present work we have developed a time dependent model of random fragmentation of molecular clouds which is based on the previous works (Feller \cite{Feller1980}; Elmegreen and Mathieu \cite{Elmegreen1983};  Chattopadhyay et al.  \cite{Chattopadhyay2003}; Chattopadhyay et al. \cite{Chattopadhyay2011}) but including turbulence of the fragments arising out of driving mechanisms due to shock waves as a result of explosion in the central region of massive galaxies and other mechanisms e.g.  supernova explosions etc. Here a hierarchical fragmentation model has been considered where number of fragments, inter occurrence time of two successive fragmentations, turbulent velocities of the fragments, are all random variables. We have considered two different distributions of turbulent velocity of the fragments, namely, a Gaussian distribution, which is a symmetric distribution and later a Gamma distribution which is merely a deviation from a symmetric distribution on the basis of two physical mechanisms. When the scale of the driving mechanism is weak, then the velocity is random in its highest mode i.e. the entropy is maximum. Then a Gaussian distribution is suitable for such physical environment. On the contrary when there is a large scale driving mechanism e.g. Galactic shocks, as a result of a big central explosion at its centre or supernova explosions of many massive stars at a time etc. then there is a large deviation from the randomness i.e. the randomness is reduced somewhat and we can consider non Gaussian distributions like  Gamma distribution having an entropy which is therefore not a maximum one (Yamamoto and Kambe \cite{yamamoto1991}; Llewellyn Smith and Gille \cite{Llewellyn1998}; Pereira et al. \cite{Pereira2016}; Jeffrey et al. \cite{Jeffrey2017}). Hence we have  considered these two distributions for the present study.
 Finally we have computed the mass spectrum in these molecular clouds of various masses and fitted segmented power laws and compared with the observed ones. The following features have been observed.\\
 \begin{itemize}
 \item For Gaussian distribution of turbulent velocities of the fragments and for the same parent cloud mass, as the efficiency ($\epsilon$) increases, the power laws become flatter in the high mass range for both high and low temperatures. In low mass range it is steeper for low temperature having both low and high efficiencies. In the low mass range it is also steeper for high temperature, low efficiency  and flatter for high temperature and high efficiency, for clouds resulting into masses comparable to those of open clusters (Tables \ref{tab:2} and \ref{tab:3}). This might be due to the fact that as efficiency increases formation of more massive stars become more frequent making the power law flatter in the high mass range contrary to the reduction of too many low mass fragments turning it to a steeper one. As the temperature increases, for higher efficiency, may be the turbulence is the key parameter which helps to produce low mass fragments in larger numbers perhaps due to elastic collisions among the fragments. Tables \ref{tab:4} - \ref{tab:5} show the estimates of the slopes (viz. $\hat{\alpha_1}, \hat{\alpha_2}$) computed with the model developed by Chattopadhyay et al. (\cite{Chattopadhyay2015}) as a robustness check. These are also consistent with the slopes found using the present model for similar values of the parameters. \\
 \item The change in the slopes (viz. ${\alpha_1}, {\alpha_2}$), for clouds leading to open clusters is much higher
 compared to those leading to globular clusters for a change of efficiency as well as change of the initial temperature of the clouds. Also the values of $\alpha_2$ are closer to Salpeter index ($\sim 2.35$) in case of clouds forming open clusters than those forming globular clusters. The values of $\alpha_1$ are much higher than the observed ones ($\sim 0.25 $-$ 0.3)$ in case of open clusters than globular clusters. This leads to the conclusion that formation of massive stars are less likely in open clusters than in globular clusters in general. This is very natural as the masses of globular clusters are largest by almost an order of two in extreme cases. This is to be noted that the turbulent structure in the present model is Gaussian. The mean value of the turbulence is proportional to the efficient mass of the parent cloud consumed in star formation. As the efficient mass increases (e.g. in case of massive clouds) the mean value of turbulent velocity also increases which is responsible for increasing the mass of the fragment. This in turn produces more massive stars resulting a rather flat mass spectrum in massive star clusters. This fact is in contrast with the case of parent clouds leading to the formation of open clusters of smaller masses.  \\

\item It is evident from Figures \ref{fig:2} - \ref{fig:4} and    Tables \ref{tab:8} - \ref{tab:9}, that with the increase of velocity dispersion ($\sim$ $\sigma$), the resulting mass spectrum has a wider range. This result might be due to the increase of collision rate among the fragments (Field and Saslaw \cite{Field1965}, equation III.1) which is responsible for producing a wider range of masses as a result of fragmentation. 
 \\
 
 \item When the turbulent distribution is one of Gamma type (i.e. positively skewed) then the mass functions both for high and low mass ranges of the fragments are comparable to Salpeter slope for high mass end and to the observations ($\sim 0.3$) of low mass stars for various combinations of initial values of the parameters (see Tables ~\ref{tab:10} - \ref{tab:11}). Several authors have pointed out that turbulent velocity deviates from Gaussian distribution (Ossenkopf and Maclow \cite{Ossenkopf2002};Kritsuk et al. \cite{Kritsuk2007}; Ragot \cite{Ragot2009}; Federrath et al. \cite{Federrath2010}; Wilczek et al. \cite{Wilczek2011}; Hennebell and Falgarone \cite{Hennebelle2012}; Krumholz and Burkhert \cite{Krumholz2016}; {Wilczek et al. \cite{Wilczek2017}}) due to large scale driving process e.g. shocks produced by supernova explosions ({Calzavara and Matzner \cite{Calzavara2004}}; Zhang and Chevalier \cite{Zhang2018}; {Sandoval et al. \cite{Sandoval2019}}) and other mechanisms e.g. gigantic explosions occurring at the centre of giant galaxies (Miesch and Bally \cite{Miesch1994}; {Sofue (\cite{Sofue1994}, \cite{Sofue2019})}; Mondal and Chattopadhyay \cite{Mondal2019}). Thus, Gaussian profile of turbulence may be associated with small scale driving mechanism e.g. coalescence and disintegration of fragments (Silk \cite{Silk1966}; {Silk and Takahashi \cite{Silk1979}; Bonnell et al. \cite{Bonnel1998}; Bonnell and Bate \cite{Bonnel2002}}), {infall of small structures} {(Takizawa \cite{Takizawa2005})}, {Active Galactic Nuclei (AGN) Jets} {(Scannapieco and Br{\"u}ggen \cite{Scannapieco2008})}, {Galaxy wakes} {(Roland \cite{Roland1981}; Bregman and David \cite{Bregman1989}; Kim \cite{Kim2007})} etc. Hence large scale driving mechanism is responsible for producing open and globular clusters those have mass functions similar to Salpeter type along with a Gamma type turbulence profile but other small scale phenomena are primarily responsible for producing open clusters having Salpeter slope with Gaussian turbulence profile. Gaussian turbulence profile may produce globular clusters in massive clouds but their mass functions will be shallower and vary much from Salpeter mass function ({Figer et al. \cite{Figer1999}; Stolte et al. \cite{Stolte2002}; Sung et al. \cite{Sung2004}; Kim et al. \cite{Kim2006}; Stolte et al. \cite{Stolte2006}; Harayama et al. \cite{Harayama2008}; Espinoza et al. \cite{Espinoza2009}}). This may be due to the fact that large scale driving mechanism accentuates the escape process of gas from parent clouds leaving little gas for formation of massive stars further which makes the mass function steeper (e.g. Salpeter type). Thus IMFs in star clusters are not universal but vary from cloud to cloud and from galaxy to galaxy ({Dabringhausen et al. \cite{Dabringhausen2009}; Gunawardhana  et al. \cite{Gunawardhana2011}; Dabringhausen et al. \cite{Dabringhausen2012}; Marks et al. \cite{Marks2012}; Ferreras et al. \cite{Ferreras2013}; Romano et al. \cite{Romano2017}}) depending on the local or global environments of the parent clouds. 
\end{itemize}

\section{CONCLUSION}  \label{CONCLUSION}

The present model deals with the turbulent structure of fragments resulting out of time dependent hierarchical fragmentation of molecular clouds lying in the central region of our Galaxy. Two patterns of turbulence have been considered, namely, Gaussian and Gamma distributions. It is found that for a Gaussian distribution, the IMFs are much flatter than Salpeter IMF in the high mass regime and steeper in the low mass regime compared to the observed slopes . On the contrary, the IMFs are comparable to the Salpeter mass function in the high mass regime for a Gamma distribution. The matching is more pronounced in case of initial higher temperature, higher efficiency and for massive parent clouds. The fact may be explained by the driving mechanism prevalent in the environment which originates at the centre of giant galaxies as a result of big central explosion and deviates the pattern of turbulence from a symmetric to one of skewed distributions.
\clearpage


\clearpage

\begin{table}
\renewcommand{\arraystretch}{1.1}
\caption{Segmented power-law fit for different initial choices of parameters (leading to open  clusters) and for a Gaussian distribution of turbulent velocity} 
\centering

\begin{tabular}{|c| c| c| c| c| c| c| c| c| c| c|}
\hline 
 $M$ & $T$   & $\epsilon$   &  $v$  &  $\eta$ & $\sigma$ & $N$  &  $\Gamma_1$  & $\alpha_1$ & $\Gamma_2$ & $\alpha_2$ \\
 (10$^4$ $M_\odot$) & (K) &  & (kms$^{-1})$  & (cm$^{-3}$)  & 
 (kms$^{-1})$ &     &      &   &              &       \\
\hline

 0.1 & 10 & 0.1 & 1.0686 & 5$\times$10$^5$  & 0.3  & 77 & 0.4627 & 0.5373 & $-$0.5343 & 1.5343  \\ 

 0.1 & 10 & 0.3 & 1.3097 & 5$\times$10$^5$ & 0.3 & 205 & 0.4014 & 0.5986 & $-$1.2822 & 2.2822  \\
  
 0.1 & 20 & 0.1 & 1.0362 & 5$\times$10$^5$  & 0.3 & 65 & 0.3543 & 0.6457 & $-$0.8725 & 1.8725  \\ 
 
 0.1 & 20 & 0.3 & 1.3441 & 5$\times$10$^5$ & 0.3 & 99 & 0.1903 & 0.8097 & $-$1.1106 & 2.1106  \\
  
 0.5 & 10 & 0.1 & 1.4495  & 10$^5$  & 0.3 & 226 & 0.4809 & 0.5110 & $-$0.8793 & 1.8793   \\ 
 
 0.5 & 10 & 0.3 & 1.8169 & 10$^5$ & 0.3 & 643 & 0.1680 & 0.8320 & $-$1.0321 & 2.0321  \\
 
 0.5 & 20 & 0.1 & 1.4171 & 10$^5$ & 0.3 & 178 & 0.3631 & 0.6369 & $-$1.1553 & 2.1553  \\ 
 
 0.5 & 20 & 0.3 & 1.7803 & 10$^5$ & 0.3 & 436 & 0.1098 & 0.8902 & $-$1.1539 & 2.1539  \\
 
  1 & 10 & 0.1 & 1.7168 & 10$^5$ & 0.3 & 388 & 0.1909 & 0.8091 & $-$1.2197 & 2.2197  \\ 
 
  1 & 10 & 0.3 & 2.1389 & 10$^5$ & 0.3 & 970 & 0.6296 & 0.3704 & $-$0.9832 & 1.9832  \\ 
 
  1 & 20 & 0.1 & 1.7529 &  10$^5$ & 0.3 & 359 & 0.6742 & 0.3258 & $-$0.9100 & 1.9100  \\ 
 
 1 & 20 & 0.3 & 2.1002 & 10$^5$ & 0.3 & 958 & 0.2097 & 0.7903 & $-$1.3906 & 2.3906  \\
 \hline 
\end{tabular}\label{tab:2}
\end{table}

\begin{table}
\renewcommand{\arraystretch}{1.1}
\caption{Segmented power-law fit for different initial choices of parameters (leading to globular  clusters), for a Gaussian distribution of turbulent velocity } 
\centering

\begin{tabular}{|c| c| c| c| c| c| c| c| c| c| c|}
\hline 
$ M$ & $T$   & $\epsilon$   &  $v$  &  $\eta$ & $\sigma$ & $N$  &  $\Gamma_1$  & $\alpha_1$ & $\Gamma_2$ & $\alpha_2$ \\
 (10$^4M_\odot$) & (K) &  & (kms$^{-1})$  & (cm$^{-3}$)  & 
 (kms$^{-1})$ &     &      &   &              &       \\
\hline

 5 & 10 & 0.1 & 2.3358 & 5$\times$10$^4$ & 0.3 & 978 & 0.4688 & 0.5312 & $-$0.9345  & 1.9345  \\ 
   
 5 & 10 & 0.3 & 2.8744 & 5$\times$10$^4$ & 0.3 & 3194 & 0.3160 & 0.6840 & $-$0.7797 & 1.7797  \\
 
 5 & 20 & 0.1 & 2.2489 & 5$\times$10$^4$ & 0.3 & 861 & 0.6026 & 0.3974 & $-$0.7654 & 1.7654  \\ 
 
 5 & 20 & 0.3 & 2.8209 & 5$\times$10$^4$ & 0.3 & 2848 & 0.8081 & 0.1919 & $-$0.5985 & 1.5985  \\
 
 10 & 10 & 0.1 & 2.6474 & 10$^4$ & 0.3 & 1461 & 0.6938 & 0.3062 & $-$0.8528 & 1.8528  \\ 
 
 10 & 10 & 0.3 & 3.2764 & 10$^4$ & 0.3 & 7384 & 0.2397 & 0.7603 & $-$0.6132 & 1.6132  \\ 
 
 10 & 20 & 0.1 & 2.6526 & 10$^4$ & 0.3 & 1485 & 0.5547 & 0.4453 & $-$1.1507 & 2.1507  \\ 
 
 10 & 20 & 0.3 & 3.3514 & 10$^4$ & 0.3 & 7143 & 0.1254 & 0.8746 & $-$0.8532 & 1.8532  \\
 
 50 & 10 & 0.1 & 3.8446 & 10$^4$ & 0.3 & 6521 & 0.7824 & 0.2176 & $-$1.7655 & 2.7655  \\ 
 
 50 & 10 & 0.3 & 4.5638 & 10$^4$ & 0.3 & 34327 & 0.1188 & 0.8812 & $-$0.3493 & 1.3493  \\
 
 50 & 20 & 0.1 & 3.7036 & 10$^4$ & 0.3 & 5701 & 0.9192 & 0.0808 & $-$1.5410 & 2.5410  \\ 
 
 50 & 20 & 0.3 & 4.5539 & 10$^4$ & 0.3 & 29534 & 0.3807 & 0.6193 & $-$1.0621 & 2.0621  \\
 
 100 & 10 & 0.1 & 4.2023 & 10$^4$ & 0.3 & 11058 & 0.8870 & 0.1130 & $-$1.6038 & 2.6038  \\ 
   
 100 & 10 & 0.3 & 5.2265 & 10$^4$ & 0.3 & 62176 & 0.1260 & 0.8740 & $-$0.9922 & 1.9922  \\ 
 
 100 & 20 & 0.1 & 4.1721 & 10$^4$ & 0.3 & 10586 & 0.8825 & 0.1175 & $-$1.9617 & 2.9617  \\ 
 
 100 & 20 & 0.3 & 5.2485 & 10$^4$ & 0.3 & 56110 & 0.2862 & 0.7138 & $-$0.5545 & 1.5545  \\
 
 \hline 
\end{tabular}\label{tab:3}
\end{table}

\clearpage

\begin{table}
\renewcommand{\arraystretch}{1}
\addtolength{\tabcolsep}{-3.5pt}
\caption{Estimates for segmented power law indices using truncated Pareto distribution (leading to open clusters), for a Gaussian distribution of turbulent velocity}

\begin{tabular}{|c |c| c| c| c| c| c| c|}
\hline 
 $M$ & $T$   & $\epsilon$ & $m_{min}$ & $m_c$ & $m_{max}$ & Estimate $\hat{\alpha_1}$ using equation(\ref{eq12}) & Estimate $\hat{\alpha_2}$ using equation(\ref{eq13}) \\
 (10$^4M_\odot$) & (K) &  &  ($M_\odot$)  & ($M_\odot$) & ($M_\odot$)  &  &   \\
\hline

  0.1 & 10 & 0.1 & 0.1069 & 0.4669 & 1.3670 &  0.3955 & 1.6123 \\ 

 0.1 & 10 & 0.3 & 0.1510 & 1.0010 & 3.1622 & 0.6033 & 2.1216 \\
  
 0.1 & 20 & 0.1 & 0.1145 & 0.4144 & 1.9123 & 0.6022  & 1.4896 \\ 
 
 0.1 & 20 & 0.3 & 0.4869 & 1.7100 & 5.9799 & 0.7798  & 1.9842 \\
  
 0.5 & 10 & 0.1 & 0.1242 & 0.8242 & 2.8242 & 0.4593  & 1.8559 \\ 
 
 0.5  & 10 & 0.3 &  0.2039 & 1.3999 & 4.4596 & 0.8941 & 2.1564  \\
 
 0.5 & 20 & 0.1 & 0.2613 & 1.8613 & 4.0613 & 0.5050  & 1.9965 \\ 
 
 0.5 & 20 & 0.3 & 0.5940 & 1.5003 & 4.2198 & 0.8998 & 2.1845 \\
 
 1 & 10 & 0.1 & 0.2106 & 1.2107 & 4.2107 &  0.8137 & 2.0369 \\ 
 
 1 & 10 & 0.3 & 0.3570 & 1.0999 & 3.7402 & 0.3654 & 1.9347 \\ 
 
 1 & 20 & 0.1 & 0.2872 & 1.4873 & 5.0873 &  0.3321 & 1.9087 \\ 
 
 1 & 20 & 0.3 & 0.4162 & 2.2509 & 5.9306 & 0.7996 & 2.4078 \\

 \hline 
\end{tabular}\label{tab:4}
\end{table}

\begin{table}
\renewcommand{\arraystretch}{1}
\addtolength{\tabcolsep}{-3.5pt}
\caption{Estimates for segmented power law indices using truncated Pareto distribution (leading to globular clusters), for a Gaussian distribution of turbulent velocity} 
\centering

\begin{tabular}{|c| c| c| c| c| c| c| c|}
\hline 
M & T   & $\epsilon$ & $m_{min}$ & $m_c$  & $m_{max}$ & Estimate $\hat{\alpha_1}$ from equation(\ref{eq12}) & Estimate $\hat{\alpha_2}$ from equation(\ref{eq13}) \\
 (10$^4M_\odot$) & (K) &  &  ($M_\odot$)  & ($M_\odot$) & ($M_\odot$)  &   &  \\
\hline

 5 & 10 & 0.1 & 0.2741 & 2.0241 & 6.0241  &  0.4215 & 2.0069  \\ 
   
 5 & 10 & 0.3 & 0.6290 & 1.7701 & 4.1495 & 0.7063 & 1.9989 \\
 
 5 & 20 & 0.1 & 0.5267 & 2.5268 & 5.5268  & 0.5825  & 1.5926  \\ 
 
 5 & 20 & 0.3 & 0.4370 & 1.6998 & 4.9102 & 0.2125 & 1.4657 \\
 
 10 & 10 & 0.1 & 0.1579 & 1.5677 & 6.3095  & 0.6293  & 2.0154  \\

 10 & 10 & 0.3 & 0.3719  & 1.3399 & 3.9003 & 0.7588 &  1.6544 \\ 
 
 10 & 20 & 0.1 & 0.2529 & 2.8028 & 7.0235  & 0.4429  & 2.1985  \\ 
 
 10 & 20 & 0.3 & 0.4559 & 1.7100 & 4.5603 & 0.7599 & 1.8049 \\
 
 50 & 10 & 0.1 & 0.5288 & 2.0290 & 6.0441  &  0.3276 & 2.4795  \\ 
 
 50 & 10 & 0.3 & 0.2119 & 1.4893 & 3.6897 & 0.7648 & 1.4578 \\
 
 50 & 20 & 0.1 & 0.5455 & 4.5023 & 8.5323 & 0.2155 & 2.2666  \\ 
 
 50 & 20 & 0.3 & 0.5689 & 1.7575 & 7.6594 & 0.6022 & 2.0001 \\
 
 100 & 10 & 0.1 & 0.5141 & 4.5143 & 8.4926  & 0.1049  & 2.5491  \\ 
   
 100 & 10 & 0.3 & 0.3800 & 2.4700 & 7.3994 & 0.8269 & 1.8999  \\ 
 
 100 & 20 & 0.1 & 0.5046 & 1.5046 &  4.3246 & 0.4569  & 1.7445  \\ 
 
 100 & 20 & 0.3 & 0.5610 & 1.9601 & 5.1903 &  0.7445  & 1.6987 \\

 \hline 
\end{tabular}\label{tab:5}
\end{table}

\clearpage

\begin{table}
\renewcommand{\arraystretch}{1.1}
\caption{Segmented power-law fit for different initial choices of parameters (leading to open clusters) and for a Gaussian distribution of turbulent velocity}
\centering

\begin{tabular}{|c| c| c| c| c| c| c| c| c| c| c|}
\hline 
 $M$ & $T$   & $\epsilon$   &  $v$  &  $\eta$ & $\sigma$ & $N$  &  $\Gamma_1$  & $\alpha_1$ & $\Gamma_2$ & $\alpha_2$ \\
 (10$^4$ $M_\odot$) & (K) &  & (kms$^{-1})$  & (cm$^{-3}$)  & 
 (kms$^{-1})$ &     &      &   &              &       \\
\hline

 0.1 & 10 & 0.1 & 0.8347 & 5$\times$10$^5$ & 1.5 & 54 & 0.6325 & 0.3675 & $-$0.4312 & 1.4312 \\ 

 0.1 & 10 & 0.3 & 0.9146   & 5$\times$10$^5$ & 1.5 & 208 & 0.8823 & 0.1177 & $-$1.0572 & 2.0572 \\
  
 0.1 & 20 & 0.1 & 1.4733 & 5$\times$10$^5$ & 1.5 & 66 & 0.8127 & 0.1873 & $-$0.5441 & 1.5441 \\ 
 
 0.1 & 20 & 0.3 & 1.3349 & 5$\times$10$^5$ & 1.5 & 163 & 0.6527 & 0.3473 & $-$1.5456 & 2.5456  \\
  
 0.5 & 10 & 0.1 & 1.5508 & 10$^5$ & 1.5 & 185 & 0.7302 & 0.2698 & $-$1.1995 & 2.1995 \\ 
 
 0.5 & 10 & 0.3 & 1.8398 & 10$^5$ & 1.5 & 565 & 0.6623 & 0.3377 & $-$1.0011 & 2.0011  \\
 
 0.5 & 20 & 0.1 & 0.8338 & 10$^5$ & 1.5 & 197 & 0.8956 & 0.1044 & $-$1.0780 & 2.0780 \\ 
 
 0.5 & 20 & 0.3 & 1.7181 & 10$^5$ & 1.5 & 419 & 0.5905 & 0.4095 & $-$0.7385 & 1.7385  \\
 
 1 & 10 & 0.1 & 1.8454 & 10$^5$ & 1.5 & 331 & 0.5100 & 0.4900 & $-$0.9615 & 1.9615   \\ 
 
  1 & 10 & 0.3 & 2.0686 & 10$^5$ & 1.5 & 534 & 0.5200 & 0.4800 & $-$1.1169 & 2.1169  \\ 
 
  1 & 20 & 0.1 & 1.6534 & 10$^5$ & 1.5 & 366 & 0.7733 & 0.2267 & $-$1.8515 & 2.8515 \\ 
 
 1 & 20 & 0.3 & 1.8822 & 10$^5$ & 1.5 & 765 & 0.2074 & 0.7926 & $-$0.7858 & 1.7858  \\

 \hline 
\end{tabular}\label{tab:6}
\end{table}

\begin{table}
\renewcommand{\arraystretch}{1}

\caption{Segmented power-law fit for different initial choices of parameters (leading to globular clusters), for a Gaussian distribution of turbulent velocity}

\begin{tabular}{|c| c| c| c| c| c| c| c| c| c| c|}
\hline 
$ M$ & $T$   & $\epsilon$   &  $v$  &  $\eta$ & $\sigma$ & $N$  &  $\Gamma_1$  & $\alpha_1$ & $\Gamma_2$ & $\alpha_2$ \\
 (10$^4$ $M_\odot$) & (K) &  & (kms$^{-1})$  & (cm$^{-3}$)  & 
 (kms$^{-1})$ &     &      &   &              &       \\
\hline

 5 & 10 & 0.1 & 2.8604 & 5$\times$10$^4$ & 2.4 & 1161 & 0.5562 & 0.4438 & $-$0.8396 & 1.8396  \\  
   
 5 & 10 & 0.3 & 2.8369 & 5$\times$10$^4$ & 2.4 & 2683 & 0.2820 & 0.7180 & $-$1.0167 & 2.0167  \\
 
 5 & 20 & 0.1 & 2.1560 & 5$\times$10$^4$ & 2.4 & 1211 & 0.5370 & 0.4630 & $-$1.0423 & 2.0423  \\ 
 
 5 & 20 & 0.3 & 2.6290 & 5$\times$10$^4$ & 2.4 & 2838 & 0.2378 & 0.7622 & $-$1.3777 & 2.3777  \\
 
 10 & 10 & 0.1 & 3.0314 & 10$^4$ & 2.4 & 2003 & 0.3365 & 0.6635 & $-$0.9794 & 1.9794  \\ 
 
 10 & 10 & 0.3 & 3.3917 & 10$^4$ & 2.4 & 5352 & 0.4618 & 0.5382 & $-$1.0622 & 2.0622  \\ 
 
 10 & 20 & 0.1 & 2.8179 & 10$^4$ & 2.4 & 1754 & 0.7214 & 0.2786 & $-$1.2564 & 2.2564  \\ 
 
 10 & 20 & 0.3 & 3.2501 & 10$^4$ & 2.4 & 5616 & 0.3939 & 0.6061 & $-$1.3498 & 2.3498  \\
 
 50 & 10 & 0.1 & 3.9735 & 10$^4$ & 3.0 & 6558 & 0.3020 & 0.6980 & $-$1.1939 & 2.1939  \\ 
 
 50 & 10 & 0.3 & 4.5977 & 10$^4$ & 3.0 & 30943 & 0.1289 & 0.8711 & $-$1.3168 & 2.3168  \\
 
 50 & 20 & 0.1 & 4.3063 & 10$^4$ & 3.0 & 6992 & 0.7727 & 0.2273 & $-$1.4041 & 2.4041  \\ 
 
 50 & 20 & 0.3 & 4.2339 & 10$^4$ & 3.0 & 28837 & 0.3706 & 0.6294 & $-$1.2364 & 2.2364  \\
 
 100 & 10 & 0.1 & 4.3928 & 10$^4$ & 3.0 & 13508 & 0.5877 & 0.4123 & $-$1.7727 & 2.7727  \\ 
   
 100 & 10 & 0.3 & 5.5400 & 10$^4$ & 3.0 & 66255 & 0.2787 & 0.7213 & $-$1.4549 & 2.4549  \\ 
 
 100 & 20 & 0.1 & 4.1474 & 10$^4$ & 3.0 & 11397 & 0.7308 & 0.2692 & $-$1.8810 & 2.8810  \\ 
 
 100 & 20 & 0.3 & 4.5922 & 10$^4$ & 3.0 & 53437 & 0.3173 & 0.6827 & $-$1.3455 & 2.3455  \\

 \hline 
\end{tabular}\label{tab:7}
\end{table}

\begin{table}
\renewcommand{\arraystretch}{1}
\addtolength{\tabcolsep}{-3.5pt}
\caption{Estimates for segmented power law indices using truncated Pareto distribution (leading to open clusters), for a Gaussian distribution of turbulent velocity}

\begin{tabular}{|c| c| c| c| c| c| c| c|}
\hline 
 $M$ & $T$   & $\epsilon$ & $m_{min}$ & $m_c$ & $m_{max}$ & Estimate $\hat{\alpha_1}$ using equation(\ref{eq12}) & Estimate $\hat{\alpha_2}$ using equation(\ref{eq13}) \\
 (10$^4$ $M_\odot$) & (K) &  &  ($M_\odot$)  & ($M_\odot$) & ($M_\odot$)  &  &   \\
\hline

  0.1 & 10 & 0.1 & 0.1245  & 0.3725 & 2.0125 &  0.3041  & 1.4229 \\ 

 0.1 & 10 & 0.3 & 0.1217 & 0.7907 & 2.6218 & 0.2054  & 1.9744 \\
  
 0.1 & 20 & 0.1 & 0.1699 & 0.4701 & 2.4299 & 0.2597  & 1.5449 \\ 
 
 0.1 & 20 & 0.3 & 0.2314 & 1.2314 & 3.8317 & 0.3764  & 2.4897 \\
  
 0.5 & 10 & 0.1 & 0.1038 & 0.3038 & 1.4038 & 0.3009  & 2.1184 \\ 
 
 0.5  & 10 & 0.3 &  0.2997 & 1.6237 & 4.8977 & 0.3296 & 2.1010  \\
 
 0.5 & 20 & 0.1 & 0.1862 & 0.6363 & 3.0359 & 0.2066  & 1.8746 \\ 
 
 0.5 & 20 & 0.3 & 0.2180 & 1.6180 & 4.8183 & 0.4115 & 1.7349 \\
 
  1 & 10 & 0.1 & 0.3270 & 1.1281 & 3.7282 & 0.5039  & 1.6197 \\ 
 
  1 & 10 & 0.3 & 0.2609 & 1.9952 & 5.2609 & 0.3926 & 2.0087 \\ 
 
  1 & 20 & 0.1 & 0.2427 & 0.9926 & 3.9930 & 0.2311  & 2.8841 \\ 
 
 1 & 20 & 0.3 & 0.2311 & 1.5488 & 4.2305 & 0.8080  & 1.6452 \\

 \hline 
\end{tabular}\label{tab:8}
\end{table}

\begin{table}
\renewcommand{\arraystretch}{1}
\addtolength{\tabcolsep}{-3.5pt}
\caption{Estimates for segmented power law indices using truncated Pareto distribution (leading to globular clusters), for a Gaussian distribution of turbulent velocity} 
\centering

\begin{tabular}{|c| c| c| c| c| c| c| c|}
\hline 
M & T   & $\epsilon$ & $m_{min}$ & $m_c$  & $m_{max}$ & Estimate $\hat{\alpha_1}$ from equation(\ref{eq12}) & Estimate $\hat{\alpha_2}$ from equation(\ref{eq13}) \\
 (10$^4$ $M_\odot$) & (K) &  &  ($M_\odot$)  & ($M_\odot$) & ($M_\odot$)  &   &  \\
\hline

 5 & 10 & 0.1 & 0.1719 & 0.9219 & 4.3722  & 0.4458  & 1.8013  \\
   
 5 & 10 & 0.3 & 0.4041 & 2.4043 & 7.2041 & 0.7243 & 2.1145 \\
 
 5 & 20 & 0.1 & 0.2455 & 0.9454 & 3.0453  & 0.3971  &  2.0421 \\
 
 5 & 20 & 0.3 & 0.3326 & 2.4547 & 6.3326 & 0.7624 & 2.3269 \\
 
 10 & 10 & 0.1 & 0.1103 & 2.1101 & 5.6094  & 0.5946  &  1.9332 \\
 
 10 & 10 & 0.3 & 0.4435  & 2.4434 & 7.2427 & 0.4982 & 2.0358  \\ 
 
 10 & 20 & 0.1 & 0.3981 & 1.2181 & 5.0118  & 0.7202  & 2.2128  \\
 
 10 & 20 & 0.3 & 0.3785 & 1.9512 & 4.5485 & 0.6063 & 2.3941 \\
 
 50 & 10 & 0.1 & 0.3051 & 2.3052 & 5.6050  & 0.5647  &  2.1865 \\ 
 
 50 & 10 & 0.3 & 0.3754 & 2.9512 & 9.9754 & 0.8797 & 2.2388 \\
 
 50 & 20 & 0.1 & 0.5143 & 3.0143 & 7.5143  &  0.2020 & 2.3142  \\
 
 50 & 20 & 0.3 & 0.4555 & 3.1622 & 9.2595 & 0.6332 & 2.2719 \\
 
 100 & 10 & 0.1 & 0.3086 & 2.3086 &  6.9090 & 0.3558  & 2.7576  \\
   
 100 & 10 & 0.3 & 0.4302 & 2.3434 & 9.1102 & 0.6299 & 2.3188  \\ 
 
  100 & 20 & 0.1 & 0.9056 & 4.0056 & 5.0060  &  0.1218 & 2.4651  \\
 
 100 & 20 & 0.3 & 0.5942 & 2.5002 & 9.9942 & 0.6193   & 2.3019  \\

 \hline 
\end{tabular}\label{tab:9}
\end{table}

\begin{table}
\renewcommand{\arraystretch}{1}

\caption{Segmented power-law fit for different initial choice of parameters (leading to open clusters) for  Gamma distribution of turbulent velocity} 

\addtolength{\tabcolsep}{-3pt}
\begin{tabular}{|c| c| c| c| c| c| c| c| c| c| c| c|}
\hline 
$ M$ & $T$   & $\epsilon$   &  $\alpha$ & $\beta$ & $v$ & $\eta$ &  N  &  $\Gamma_1$  & $\alpha_1$ & $\Gamma_2$ & $\alpha_2$ \\
 (10$^4$ $M_\odot$) & (K) &  &  &   & 
 (kms$^{-1})$ & (cm$^{-3}$)     &     &      & &        &       \\
\hline

 0.1 & 10 & 0.1 & 6.029 & 10 & 0.6029 & 5$\times$10$^5$ & 63  & 0.6521 & 0.3479 & $-$1.0224 & 2.0224  \\ 

 0.1 & 10 & 0.3 & 13.339 & 10 & 1.3339 & 5$\times$10$^5$ &  147 & 0.2859 & 0.7141 & $-$1.0143 & 2.0143  \\
  
 0.1 & 20 & 0.1 & 6.223 & 10 & 0.6223 & 5$\times$10$^5$ & 78 & 0.4527 & 0.5473 & $-$1.2149 & 2.2149  \\ 
 
 0.1 & 20 & 0.3 & 16.286 & 10 & 1.6286 & 5$\times$10$^5$ &  201 & 0.2809 & 0.7191 & $-$1.3161 & 2.3161  \\
  
 0.5 & 10 & 0.1 & 12.521 & 10 & 1.2521 & 10$^5$ & 249  & 0.6550 & 0.3450 & $-$1.0568 & 2.0568  \\ 
 
 0.5 & 10 & 0.3 & 17.816 & 10 & 1.7816 & 10$^5$ & 520 & 0.5085 & 0.4915 & $-$1.5808 & 2.5808  \\
 
 0.5 & 20 & 0.1 & 7.925 & 10 & 0.7925 & 10$^5$ &  193 & 0.6284 & 0.3716 & $-$1.6011 & 2.6011  \\ 
 
 0.5 & 20 & 0.3 & 15.497 & 10 & 1.5497 & 10$^5$ &  633 & 0.2859 & 0.7141 & $-$1.5710 & 2.5710  \\
 
 1 & 10 & 0.1 & 12.458 & 10 & 1.2458 & 10$^5$ & 522  & 0.7488 & 0.2512 & $-$1.6801 & 2.6801  \\ 
 
  1 & 10 & 0.3 & 20.362 & 10 & 2.0362 & 10$^5$ & 1115 & 0.1464 & 0.8536 & $-$1.5438 & 2.5438  \\ 
 
 1 & 20 & 0.1 & 18.216 & 10 & 1.8216 & 10$^5$ & 395  & 0.4510 & 0.5490 & $-$1.1651 & 2.1651  \\ 
 
 1 & 20 & 0.3 & 21.954 & 10 & 2.1954 & 10$^5$ &  952 & 0.5630 & 0.4370 & $-$1.3916 & 2.3916  \\

 \hline
\end{tabular}\label{tab:10}
\end{table}

\begin{table}
\renewcommand{\arraystretch}{1}
\addtolength{\tabcolsep}{-3pt}
\caption{Segmented power law fit for different initial choice of parameters (leading to globular clusters) for Gamma distribution of turbulent velocity} 
\centering

\begin{tabular}{|c| c| c| c| c| c| c| c| c| c| c| c|}
\hline 
 $M$ & $T$   & $\epsilon$   & $\alpha$  & $\beta$ &  $v$  &  $\eta$ & N  &  $\Gamma_1$  & $\alpha_1$ & $\Gamma_2$ & $\alpha_2$ \\
 (10$^4$ $M_\odot$) & (K) &  &   &   & 
 (kms$^{-1})$ &   (cm$^{-3}$)  &      &     &  &         &       \\
\hline

  5 & 10 & 0.1 & 27.621 & 10 & 2.7621 & 5$\times$10$^4$ & 1004  & 0.5677 & 0.4323 & $-$1.2486 &  2.2486  \\ 
   
 5 & 10 & 0.3 & 17.572 & 10 & 1.7572& 5$\times$10$^4$ &  4816 & 0.4993 & 0.5007 & $-$1.1989 & 2.1989  \\
 
 5 & 20 & 0.1 & 25.854 & 10 & 2.5854 & 5$\times$10$^4$ & 1236  & 0.5406 & 0.4594 & $-$1.4301 & 2.4301   \\ 
 
 5 & 20 & 0.3 & 17.216 & 10 & 1.7216 & 5$\times$10$^4$ &  3204 & 0.2638 & 0.7362 & $-$1.5732 & 2.5732 \\
 
 10 & 10 & 0.1 & 22.883& 10 & 2.2883 & 10$^4$ & 2110 & 0.6130  & 0.3870 & $-$1.1739 & 2.1739    \\ 
 
 10 & 10 & 0.3 & 23.753 & 10 & 2.3753 & 10$^4$ &  11488 & 0.5914 & 0.4086 & $-$1.2022 & 2.2022  \\ 
 
  10 & 20 & 0.1 & 19.004 & 10 & 1.9004 & 10$^4$ &  1937 & 0.4719 & 0.5281 & $-$1.3338 & 2.3338   \\ 
 
 10 & 20 & 0.3 & 19.392 & 10 & 1.9392 & 10$^4$ &  12717 & 0.5295 & 0.4705 & $-$1.3870 & 2.3870  \\
 
 50 & 10 & 0.1 & 38.457 & 10 & 3.8457 & 10$^4$ & 11420  & 0.8025 & 0.1975 & $-$1.5242 & 2.5242   \\ 
 
 50 & 10 & 0.3 & 31.314 & 10 & 3.1314 & 10$^4$ &  31032 & 0.5049 & 0.4951 & $-$1.0126 & 2.0126  \\
 
 50 & 20 & 0.1 & 25.761 & 10 & 2.5761 & 10$^4$ & 16967  & 0.9569 & 0.0431 & $-$1.6882 & 2.6882   \\ 
 
 50 & 20 & 0.3 & 28.470 & 10 & 2.8470 & 10$^4$ &  27209 & 0.5532 & 0.4468 & $-$1.2696 & 2.2696 \\
 
 100 & 10 & 0.1 & 36.569 & 10 & 3.6569 & 10$^4$ & 20573  & 0.7127 & 0.2873 & $-$1.5812 & 2.5812   \\ 
   
 100 & 10 & 0.3 & 44.673 & 10 & 4.4673 & 10$^4$ &  66296 & 0.2387 & 0.7613 & $-$1.4241 & 2.4241  \\ 
 
 100 & 20 & 0.1 & 47.131 & 10 & 4.7131 & 10$^4$ & 20322  & 0.9229 & 0.0771 & $-$1.9393 & 2.9393    \\ 
 
 100 & 20 & 0.3 & 31.061 & 10 & 3.1061 & 10$^4$ &  55789 & 0.6318 & 0.3682 & $-$1.3539 & 2.3539  \\

 \hline 
\end{tabular}\label{tab:11}
\end{table}

\clearpage

\begin{table}
\renewcommand{\arraystretch}{1}
\addtolength{\tabcolsep}{-3.8pt}
\caption{Estimates for segmented power law indices using truncated Pareto distribution (leading to open clusters), for Gamma distribution of turbulent velocity} 
\centering
\begin{tabular}{|c| c| c| c| c| c| c| c| c| c| c| c| c|} 
\hline 
 $M$ & $T$   & $\epsilon$ & $m_{min}$ & $m_c$ & $m_{max}$ & Estimate  $\hat{\alpha_1}$ using equation(\ref{eq12}) & Estimate $\hat{\alpha_2}$ using equation(\ref{eq13}) \\
 (10$^4$ $M_\odot$) & (K) &  &  ($M_\odot$)  & ($M_\odot$) & ($M_\odot$)  &  &   \\
\hline

 0.1 & 10 & 0.1 & 0.1778 & 0.3548 & 1.3182 & 0.3502  & 1.9553  \\ 

 0.1 & 10 & 0.3 & 0.2189 & 1.0025 & 2.7797 & 0.7049 &  2.1265\\
  
 0.1 & 20 & 0.1 & 0.1584 & 0.6309 & 2.1183 & 0.6321  & 2.2599  \\ 
 
 0.1 & 20 & 0.3 & 0.2480 & 0.8879 &  2.0897 &  0.2774 & 2.2671 \\
  
 0.5 & 10 & 0.1 & 0.1519 & 0.7519 & 2.7027 & 0.3099  & 2.0374  \\ 
 
 0.5  & 10 & 0.3 & 0.2299 & 1.6998 & 4.8194 & 0.4905 & 2.4545 \\
 
 0.5 & 20 & 0.1 & 0.5053 & 1.5055 & 3.9057 & 0.4040  & 2.5119  \\ 
 
 0.5 & 20 & 0.3 & 0.4169 & 1.8399 & 4.8797 & 0.7355 & 2.4220\\
 
 1 & 10 & 0.1 & 0.2098 & 0.7009 & 4.4096 & 0.2551  & 2.3974  \\ 

  1 & 10 & 0.3 & 0.3430 & 1.4501 & 5.2504 & 0.9022 & 2.3142 \\ 
 
 1 & 20 & 0.1 & 0.3884 & 1.5885 & 4.5885 & 0.5523  &  2.1114 \\ 
 
 1 & 20 & 0.3 & 0.6850 & 1.5599 & 5.1903 & 0.5050  & 2.3189 \\

 \hline 
\end{tabular}\label{tab:12}
\end{table}

\begin{table}
\renewcommand{\arraystretch}{1}
\addtolength{\tabcolsep}{-3.5pt}
\caption{Estimates for segmented power law indices using truncated Pareto distribution  (leading to globular clusters), for a Gamma distribution of turbulent velocity}

\begin{tabular}{|c| c| c| c| c| c| c| c|}
\hline 
$M$ & $T$   & $\epsilon$ & $m_{min}$ & $m_c$  & $m_{max}$ & Estimate $\hat{\alpha_1}$ from equation(\ref{eq12}) & Estimate $\hat{\alpha_2}$ from equation(\ref{eq13}) \\
 (10$^4$ $M_\odot$) & (K) &  &  ($M_\odot$)  & ($M_\odot$) & ($M_\odot$)  &   &  \\
\hline

  5 & 10 & 0.1 & 0.3630 & 2.7631 & 5.8632  &  0.4221 & 2.2613  \\  
   
 5 & 10 & 0.3 & 0.2210 & 1.0997 & 5.1701 & 0.4884 &  2.1006\\
 
  5 & 20 & 0.1 & 0.3126 & 1.5125 &  4.8126 & 0.3784  & 2.4447  \\ 
 
 5 & 20 & 0.3 & 0.6820 & 2.2202 & 8.3695 & 0.7495 & 2.5729 \\
 
 10 & 10 & 0.1 & 0.3004 & 2.1003 & 6.6008  & 0.3563  & 2.1465  \\ 
 
 10 & 10 & 0.3 & 0.2239 & 1.0099 & 5.0199 & 0.3997 & 2.3119  \\ 
 
  10 & 20 & 0.1 & 0.3519 & 2.1522 & 8.1526  & 0.5113  & 2.1653  \\ 
 
 10 & 20 & 0.3 & 0.2139 & 0.9590 & 4.9000 & 0.4891 & 2.3557 \\
 
  50 & 10 & 0.1 & 0.6070 & 2.1071 & 8.4062  & 0.2667  & 2.3362  \\ 
 
 50 & 10 & 0.3 & 0.3850 & 1.5399 & 6.0394 & 0.4993 & 2.1147 \\
 
 50 & 20 & 0.1 & 0.2010 & 1.4018 & 6.0010  &  0.0509 & 2.5792 \\ 
 
 50 & 20 & 0.3 & 0.5830 & 2.1399 & 8.3502 & 0.3962 & 2.3161 \\
 
 100 & 10 & 0.1 & 0.6052 & 1.8050 &  10.4954 & 0.2279  & 2.5883  \\ 
   
 100 & 10 & 0.3 & 0.5038 & 1.2589 & 9.0635 &  0.7255 & 2.4301 \\ 
 
 100 & 20 & 0.1 & 0.6006 & 2.3998 & 7.4026  & 0.2173  & 2.8181  \\ 
 
 100 & 20 & 0.3 & 0.4665 & 1.3970 & 10.2329 &  0.3651  & 2.3559 \\

 \hline 
\end{tabular}\label{tab:13}
\end{table}

\clearpage

\section{Acknowledgements}
The author S.P. is very much grateful to the Department of Science and Technology (DST),  India, for approving a  JRF grant for the work. The authors are also thankful to A.K.Chattopadhyay for useful discussions. The authors are very much thankful to the referee for valuable suggestions which have improved the work to a great extent.

\bibliographystyle{elsarticle-num}
\bibliography{Suman_ms}

\end{document}